\documentclass[num-refs,sort&compress]{wiley-article}
\usepackage{upgreek,amssymb,amsmath,epstopdf}

\papertype{FULL PAPER}

\title{LORAKI:  Autocalibrated Recurrent Neural Networks for Autoregressive MRI Reconstruction in k-Space}

\author[1]{Tae Hyung Kim}
\author[1,2]{Pratyush Garg}
\author[1]{Justin P. Haldar}

\affil[1]{Ming Hsieh Department of Electrical and Computer Engineering, University of Southern California, Los Angeles, CA, 90089, USA}
\affil[2]{Department of Electrical Engineering, Indian Institute of Technology (IIT) Kanpur, Kanpur, India}

\corraddress{Tae Hyung Kim, University of Southern California,  3710~McClintock Avenue, RTH~\#317, Los Angeles, CA, 90089}
\corremail{taehyung@usc.edu}

\fundinginfo{This work was supported in part by a USC Annenberg Fellowship, a Kwanjeong Educational Foundation Scholarship, NSF research award CCF-1350563, and NIH research awards R21-EB022951, R01-MH116173, R01-NS074980, R01-NS089212, and R33-CA225400.}

\DeclareMathOperator*{\argmin}{\arg\min}

\runningauthor{Tae Hyung Kim et al.}

\begin{document}

\maketitle

\begin{abstract}
\small 
{\bf Purpose:} To propose and evaluate a new MRI reconstruction method named LORAKI that trains an autocalibrated scan-specific recurrent neural network (RNN) to recover missing k-space data.

\noindent {\bf Methods:} Methods like GRAPPA, SPIRiT, and AC-LORAKS assume that k-space data has shift-invariant autoregressive structure, and that the scan-specific autoregression relationships needed to recover missing samples can be learned from fully-sampled autocalibration (ACS) data.  Recently, the structure of the linear GRAPPA method has been translated into a nonlinear deep learning method named RAKI. RAKI uses ACS data to train an artificial neural network to  interpolate missing k-space samples, and often outperforms GRAPPA.  In this work, we apply a similar principle to translate the linear AC-LORAKS method (simultaneously incorporating support, phase, and parallel imaging constraints) into a nonlinear deep learning method named LORAKI. 

Since AC-LORAKS is iterative and convolutional, LORAKI takes the form of a convolutional RNN.  This new architecture admits a wide range of sampling patterns, and even calibrationless patterns are possible if synthetic ACS data is generated.

The performance of LORAKI was evaluated with retrospectively undersampled brain datasets, with comparisons against other related reconstruction methods.

\noindent {\bf Results:} Results suggest that LORAKI can provide improved reconstruction compared to other scan-specific autocalibrated reconstruction methods like GRAPPA, RAKI, and AC-LORAKS.  

\noindent {\bf Conclusion:} LORAKI offers a new deep-learning approach to MRI reconstruction based on RNNs in k-space, and enables improved image quality and enhanced sampling flexibility.

\end{abstract}

\normalsize
\section{Introduction}

Slow data acquisition speed has always been one of the biggest impediments to MRI, and developing new methods to improve acquisition speed has been a major research focus for more than four decades.  One way to achieve faster data acquisition is to sample k-space data below the Nyquist rate, and then use constraints and advanced reconstruction methods to compensate for the missing information.  There is a long history of such methods, ranging  from classical constrained image reconstruction \cite{liang1992} and parallel imaging methods \cite{hamilton2017}, to more recent sparse \cite{lustig2007}, low-rank \cite{liang2007,haldar2010a,goud2011}, and structured low-rank \cite{liang1989,shin2014,haldar2014} modeling methods, to very recent machine learning methods \cite{wang2016, hammernik2018, schlemper2018, jin2017, zhu2018, han2018a,  mardani2019, aggarwal2019, akcakaya2019}.

In this work, we propose and evaluate LORAKI, a shift-invariant nonlinear autoregressive autocalibrated (AA) deep learning approach for MRI reconstruction that combines features of existing LORAKS \cite{haldar2014,haldar2016, haldar2017a} and RAKI \cite{akcakaya2019} reconstruction methods.  

RAKI is a nonlinear deep learning AA method that uses a small amount of autocalibration (ACS) data to train a scan-specific feedforward convolutional neural network (CNN) that can perform autoregressive k-space interpolation.  RAKI can be viewed as a deep learning version of the widely used non-iterative GRAPPA parallel imaging method \cite{griswold2002}.  On the other hand, Autocalibrated LORAKS (AC-LORAKS) \cite{haldar2015b,lobos2018} is a linear AA method that imposes substantially more constraints than GRAPPA (i.e., support and phase constraints in addition to parallel imaging constraints), generally outperforms GRAPPA, and can be implemented algorithmically as a series of convolution operations \cite{kim2018c,kim2018}.  

Our proposed LORAKI approach combines the nonlinear deep learning approach of RAKI with the additional constraints and iterative convolutional characteristics of AC-LORAKS.  This results in a convolutional recurrent neural network (RNN) architecture for LORAKI, which is different from the feedforward CNN architecture employed by RAKI.  We hypothesized that LORAKI would have advantages over RAKI for the same reasons that AC-LORAKS typically outperforms GRAPPA.  We also hypothesized that LORAKI could outperform AC-LORAKS because nonlinear AA methods like NL-GRAPPA \cite{chang2011}, KerNL \cite{lyu2019}, and RAKI \cite{akcakaya2019} have already been demonstrated to have performance advantages over earlier linear AA methods.

Like both AC-LORAKS and RAKI, the proposed LORAKI network is trained to be scan-specific (i.e., it should not be expected to generalize to other scans) based on a small amount of ACS data.  This is very different from the the large multi-subject databases of training data required by most other deep learning MRI reconstruction methods that attempt to train generalizable reconstruction procedures \cite{wang2016, hammernik2018, schlemper2018, jin2017, zhu2018, han2018a,  mardani2019, aggarwal2019}.  This means that ``Big Data" is not required, alleviating one of the main drawbacks of most other deep learning methods.   Like AC-LORAKS, LORAKI is also expected to be compatible with a wide range of autocalibrated Cartesian k-space sampling patterns.

Our results with retrospectively undersampled brain MRI images confirm that LORAKI can have performance advantages over both AC-LORAKS and RAKI, and that LORAKI can accommodate various sampling strategies.  A preliminary account of portions of this work will be presented in an upcoming conference \cite{kim2019}.

\section{Theory}

\subsection{Autoregressive Modeling and Autocalibration}

Many existing image reconstruction methods assume that k-space data possesses shift-invariant autoregressive structure, meaning that a missing data sample can be accurately extrapolated or interpolated based on the values of neighboring samples if autoregression relationships can be learned.  As reviewed in a recent tutorial \cite{haldar2019}, linear shift-invariant autoregressive structure can be derived in many settings as a natural consequence of limited image support \cite{haldar2014}, smooth image phase \cite{haldar2014, huang2009}, transform domain sparsity \cite{liang1989, haldar2015a,ongie2016, jin2016}, inter-image correlations in multi-contrast imaging scenarios \cite{bilgic2018}, and/or inter-channel correlations in parallel imaging contexts \cite{sodickson1997}.  AA methods are a special case of autoregressive methods, and their defining characteristic is that they learn the shift-invariant autoregression relationships from a fully-sampled region of k-space providing ACS data.  In a second step, AA methods then apply this learned information for image reconstruction.  

Most existing AA methods have assumed a linear autoregression relationship \cite{haldar2019}, including SMASH \cite{sodickson1997}, GRAPPA \cite{griswold2002},  SPIRiT \cite{lustig2010}, PRUNO \cite{zhang2011}, AC-LORAKS \cite{haldar2015b,lobos2018}, and other methods that haven't been given names or acronyms \cite{liang1989,huang2009,ongie2016}.  Lately, there has also been growing interest in nonlinear AA methods like  NL-GRAPPA \cite{chang2011}, KerNL \cite{lyu2019}, and RAKI \cite{akcakaya2019}, where the use of nonlinearity seems to improve empirical performance. 

Since GRAPPA, RAKI, and AC-LORAKS are the methods that are most relevant to our story, we provide a brief overview of these methods in the sections below, before describing our proposed LORAKI approach.  

In what follows, we assume that k-space is sampled on a Cartesian integer lattice, and use the integer vector $\mathbf{k}$ to denote the coordinates with respect to that lattice.  For example, in 2D MRI with sampling intervals of $\Delta k_x$ and $\Delta k_y$, the k-space location $(m\Delta k_x, n \Delta k_y)$ would correspond to the integer vector $\mathbf{k} = [m, n]^T$. Assuming a multi-channel experiment with a total of $L$ channels, we will use $d_\ell[\mathbf{k}]$ denote the k-space sample at lattice position $\mathbf{k}$ from coil $\ell$.

\subsubsection{GRAPPA}
GRAPPA~\cite{griswold2002} assumes that there exist shift-invariant linear interpolation relationships in k-space such that the value of $d_\ell[\mathbf{k}]$ can be accurately predicted as a linear combination of neighboring k-space samples according to
\begin{equation}
d_\ell[\mathbf{k}] \approx \sum_{c=1}^L \sum_{\mathbf{m}\in \Lambda_{\mathbf{k}}}  w_{\ell,\mathbf{m},c} d_c[\mathbf{k} - \mathbf{m}].\label{eq:GRAPPA}
\end{equation}
Here, $\Lambda_{\mathbf{k}}$ is the set of integer shift vectors that specify the relative positions of the local neighbors that will be used to interpolate point $\mathbf{k}$, and $w_{\ell,\mathbf{m},c}$ are the GRAPPA kernel weights.  Because GRAPPA assumes this relationship is shift-invariant (i.e., Eq.~\eqref{eq:GRAPPA} should be valid at every k-space position with the same kernel weights), it can be represented in convolutional form, and a small amount of ACS data can be used to train the values of the kernel weights  \cite{griswold2002}.  However, a different set of kernel weights needs to be estimated for each distinct configuration of the local sampling neighborhood.  GRAPPA is usually applied in scenarios with uniform undersampling, in which case there is a lot of repetition in the local sampling configuration, resulting in only a small number of distinct neighborhood configurations $\Lambda_{\mathbf{k}}$.  We will assume that there are only $J$ distinct values of $\Lambda_{\mathbf{k}}$, denoted by $\Lambda_j$ for $j=1,\ldots, J$.

For the sake of concreteness in what follows (and without loss of generality), we will assume a 2D imaging scenario with an $N_1 \times N_2$ grid of nominal sampling positions and a rectangular GRAPPA kernel of size $R_1 \times R_2$.   We will also assume that non-overlapping binary k-space masks $\mathbf{g}_j \in \mathbb{R}^{N_1\times N_2}$ have been constructed for $j=1,\ldots,J$, such that  $\mathbf{g}_j[m,n] = 1$ if the $(m,n)$th k-space position was unsampled and possesses local sampling configuration $\Lambda_j$, and $\mathbf{g}_j[m,n] = 0$ otherwise.  

\begin{figure}[t]
\centering
\includegraphics{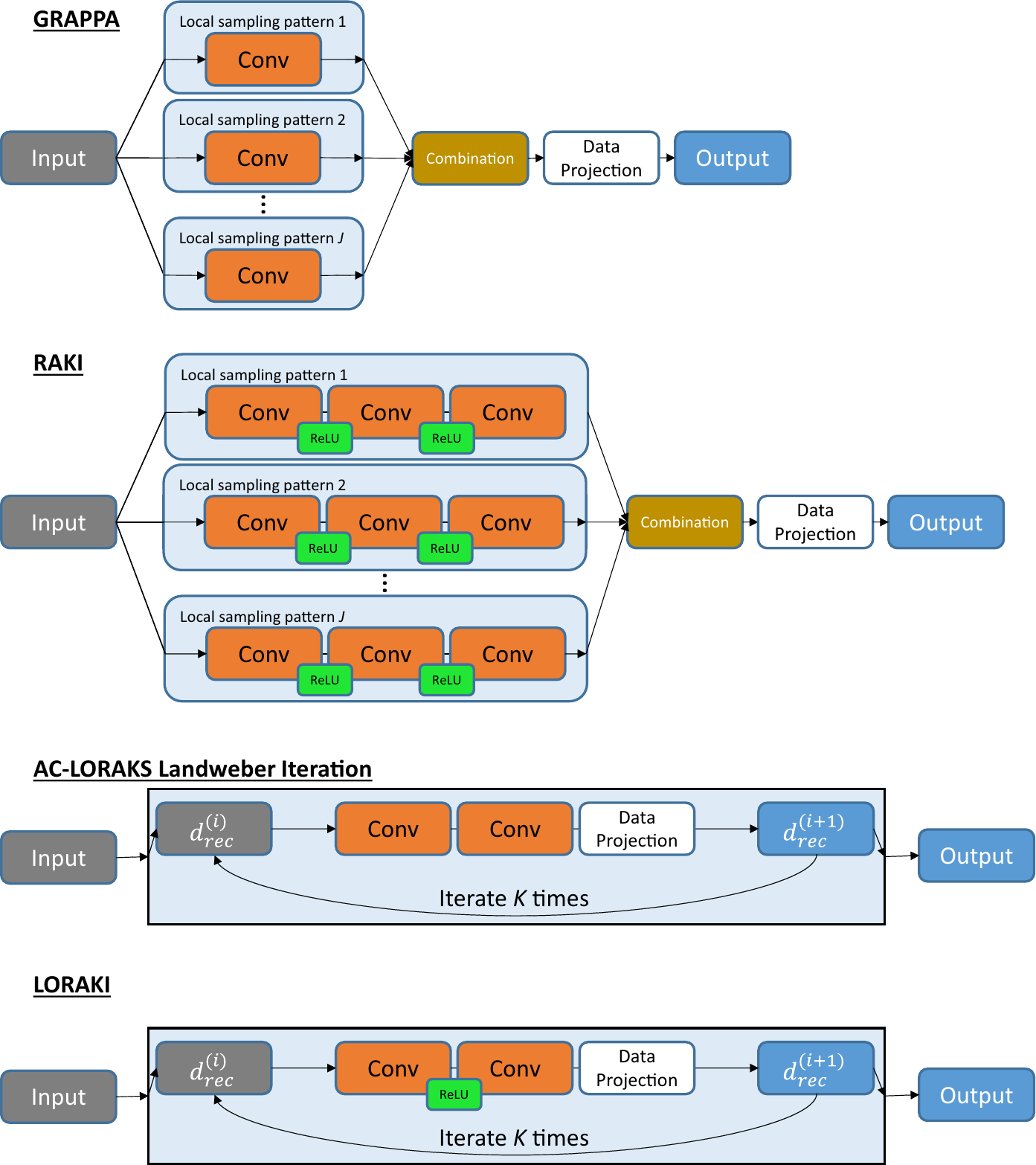}		
\caption{Neural network representations of GRAPPA, RAKI, AC-LORAKS (with Landweber iteration), and LORAKI.}
\label{fig:network_structure}
\end{figure}

GRAPPA can be equivalently viewed as a single-layer CNN without bias terms or activation functions.    In the notation of CNNs, GRAPPA is succinctly represented as 
\begin{equation}
[\mathbf{d}_{rec}]_\ell \approx [\mathbf{d}_{zp}]_\ell + \sum_{j=1}^J \mathbf{g}_j \odot [f_{\ell}(\mathbf{d}_{zp})]_j,\label{grappa_nn}
\end{equation}
for $\ell = 1,\ldots,L$.  Here, $\odot$ denotes the Hadamard product (elementwise multiplication); for arbitrary $\mathbf{a}$, $[\mathbf{a}]_\ell$ denotes the $\ell$th channel of $\mathbf{a}$; $\mathbf{d}_{rec} \in \mathbb{C}^{N_1\times N_2 \times L}$ is the output set of reconstructed k-space data; $\mathbf{d}_{zp} \in \mathbb{C}^{N_1 \times N_2 \times L}$ is the input set of acquired multi-channel data with unsampled k-space locations filled with zeros; and each $f_{\ell}(\cdot): \mathbb{C}^{N_1 \times N_2 \times L} \rightarrow \mathbb{C}^{N_1 \times N_2 \times J}$ represents the linear convolution layer corresponding to filtering the input signal with the $J$ sets of GRAPPA weights $\mathbf{w}_{\ell,j} \in \mathbb{C}^{R_1 \times R_2 \times L}$ for the $\ell$th output channel and $j$th local sampling configuration. In particular,
\begin{equation}
[f_{\ell}(\mathbf{d}_{zp})]_j = \sum_{c=1}^L [\mathbf{w}_{\ell,j}]_c \otimes [\mathbf{d}_{zp}]_c
\end{equation}
for $j=1,\ldots,J$, where $\otimes$ denotes convolution. The neural network representation for GRAPPA is illustrated in Fig.~\ref{fig:network_structure}.

The kernel weights $\mathbf{w}_{\ell,j} \in \mathbb{C}^{R_1\times R_1 \times L}$ for $\ell=1,\ldots,L$ and $j=1,\ldots,J$ in conventional GRAPPA are trained with a least-squares loss function \cite{griswold2002}, where the ACS data is subsampled according to $\Lambda_j$ to generate paired fully-sampled and undersampled training examples.  It should be noted that since the ACS data is usually smaller than $N_1 \times N_2$, the training of the $\mathbf{w}_{\ell,j}$ kernels is usually performed with adjusted input and output variable sizes (i.e., respectively replacing $N_1$ and $N_2$ with $M_1$ and $M_2$ in the above, assuming the ACS training data has size $M_1\times M_2$).

\subsubsection{RAKI}
RAKI~\cite{akcakaya2019} extends GRAPPA by using multiple convolution layers along with ReLU activation functions.  In particular, the RAKI network can be represented as
\begin{equation}
[\mathbf{d}_{rec}]_{\ell} \approx [\mathbf{d}_{zp}]_\ell + \sum_{j=1}^J \mathbf{g}_j \odot \left[f_{\ell,3}(relu(f_{\ell,2}(relu(f_{\ell,1}(\mathbf{d}_{zp})))))\right]_j. \label{raki_nn}
\end{equation}
for $\ell=1,\ldots,L$.  Note that the original RAKI formulation was described assuming simple uniform 1D undersampling along the phase encoding dimension, leading to certain differences from the more general formulation we present here.  In this expression, $f_{\ell,1}(\cdot)$, $f_{\ell,2}(\cdot)$, and $f_{\ell,3}(\cdot)$ are each linear convolution layers without bias terms, and the ReLU activation function $relu(\cdot)$ is an elementwise operation that outputs a vector with the same size as the input, with $i$th element of the output of $relu(\mathbf{x})$ defined as $\max(x_i,0)$. The structure of the RAKI CNN is also shown in Fig.~\ref{fig:network_structure}.     The nonlinear ReLU activation functions are the key features that distinguish RAKI from GRAPPA.  In particular, because of the linearity and associativity properties of convolution, applying a series of convolution layers without any nonlinearities between them is functionally equivalent to applying a single convolution layer, which would cause Eq.~\eqref{raki_nn} to effectively become an overparameterized version of Eq.~\eqref{grappa_nn} if the nonlinearities were removed.

Since the technology for complex-valued neural networks is less developed than for real-valued neural networks, RAKI was practically implemented by treating the real and imaginary components as separate real-valued channels, thereby doubling the effective number of channels \cite{akcakaya2019}.  In particular, RAKI uses $\mathbf{d}_{rec} \in \mathbb{R}^{N_1\times N_2 \times 2L }$  and $\mathbf{d}_{zp} \in \mathbb{R}^{N_1\times N_2 \times 2L}$.  Each of the convolution layers has a similar multi-channel structure to that described above for GRAPPA.  In particular, $f_{\ell,1}(\cdot)$ maps variables in $\mathbb{R}^{N_1\times N_2 \times 2L}$ to variables in $\mathbb{R}^{N_1\times N_2 \times C_1}$ by convolving its input with $C_1$ multi-channel kernels, where each kernel belongs to $\mathbb{R}^{R_{11} \times R_{21} \times 2L}$; $f_{\ell,2}(\cdot)$ maps variables in $\mathbb{R}^{N_1\times N_2 \times C_1}$ to variables in $\mathbb{R}^{N_1\times N_2 \times C_2}$ by convolving its input with $C_2$ multi-channel kernels, where each kernel belongs to $\mathbb{R}^{R_{12} \times R_{22} \times C_1}$; and $f_{\ell,3}(\cdot)$ maps variables in $\mathbb{R}^{N_1\times N_2 \times C_2}$ to variables in $\mathbb{R}^{N_1\times N_2 \times J}$ by convolving its input with $J$ multi-channel kernels, where each kernel belongs to $\mathbb{R}^{R_{13} \times R_{23} \times C_2}$.  The variables $C_1$, $C_2$ and $R_{ij}$ for $i=1,2$ and $j=1,2,3$ are all user-selected parameters. 

RAKI uses the same least-squares loss function and same training data as GRAPPA, although needs a more complicated training procedure because RAKI is nonlinear, and the simple linear least-squares techniques used by GRAPPA are not applicable.  To overcome this, RAKI can be trained using backpropagation to minimize the nonlinear least-squares loss function \cite{akcakaya2019}.  However, because RAKI generally has more parameters and is designed capture more complicated autoregressive structure than GRAPPA, it generally needs more ACS data than GRAPPA does to achieve good reconstruction results.

\subsubsection{LORAKS and AC-LORAKS}
The LORAKS framework~\cite{haldar2014,haldar2016,haldar2017a} is based on a theoretical association between autoregressive k-space structure and a variety of classical image reconstruction contraints (including limited support, smooth phase, sparsity, and parallel imaging constraints).  Specifically, LORAKS is based on the observation that when one or more of these classical constraints are satisfied by a given image, then the k-space data will approximately obey at least one (and frequently many more than one) linear autoregression relationship.  The existence of such linear autogression relationships implies that an appropriately-constructed structured matrix (e.g., convolution-structured Hankel or Toeplitz matrices) formed from the k-space data will have distinct nullspace vectors associated with each linear autoregression relationship.  This implies that such a matrix will have low-rank structure, which enables constrained image reconstruction from undersampled k-space data using modern low-rank matrix recovery methods.  A nice feature of this approach is that users of LORAKS do not need to make prior modeling assumptions about the support, phase, or parallel imaging characteristics of the images -- all of this information is implicitly captured by the nullspace of the structured matrix, which is estimated automatically from the data.  This automatic adaptation means that the LORAKS approach can still be applied in cases where the image may not obey all of the constraints that motivate the LORAKS framework.  Related structured low-rank matrix modeling approaches that have similar characteristics include Refs.~\cite{liang1989, shin2014, jin2016, ongie2016}.  

The original implementations of LORAKS were compatible with calibrationless k-space sampling \cite{haldar2014a}, and reconstructed undersampled k-space data by solving a nonconvex matrix recovery problem.  However, it was later observed that substantial  improvements in computational complexity could be achieved if ACS data were acquired, resulting in a fast linear AA method called AC-LORAKS \cite{haldar2015b}.  In particular, in the first step of AC-LORAKS, the ACS data is formed into a structured ``calibration'' matrix, and the nullspace of this calibration matrix is estimated.  Subsequently, the fully sampled k-space data is reconstructed by solving a simple linear least-squares problem:
\begin{equation}
\mathbf{d}_{rec} = \argmin_{\mathbf{d}\in \mathbb{C}^{N_1 \times N_2 \times L}} \| \mathcal{P}(\mathbf{d}) \mathbf{N} \|_F^2 \text{  s.t.  }  \mathcal{M}(\mathbf{d}) = \mathbf{d}_{zp}.\label{eq:acloraks}
\end{equation}
In this expression,  the operator $\mathcal{P}(\cdot): \mathbb{C}^{N_1 \times N_2 \times L} \rightarrow \mathbb{C}^{P \times Q}$ maps the vector of k-space data into a structured low rank matrix that is expected to have low-rank; the columns of the matrix $\mathbf{N} \in \mathbb{C}^{Q \times C}$ correspond to a collection of $C$ nullspace vectors obtained from the calibration matrix; and the linear operator $\mathcal{M}(\cdot): \mathbb{C}^{N_1 \times N_2 \times L} \rightarrow \mathbb{C}^{N_1 \times N_2 \times L}$ is a masking operator that sets k-space sample values equal to zero if they were not measured during data acquisition.  While this linear leasts-squares problem can be solved analytically in principle, the large size of the matrices in Eq.~\eqref{eq:acloraks} means that practical implementations usually rely on iterative least-squares solvers.  

Note that AC-LORAKS is strongly inspired by previous autocalibrated low-rank modeling work by Liang \cite{liang1989,liang2007}, and can also be viewed as a generalization of both the PRUNO \cite{zhang2011} and SPIRiT \cite{lustig2010} reconstruction methods.  In particular, AC-LORAKS reduces to PRUNO \cite{zhang2011} if the $\mathcal{P}(\cdot)$ operator is designed to construct a structured matrix that only imposes limited support and parallel imaging constraints  (but not the smooth phase constraints that are available through novel LORAKS matrix constructions), and PRUNO reduces to the previous SPIRiT technique \cite{lustig2010} if the number of nullspace vectors is set to $C=1$.

An open source implementation of AC-LORAKS that relies on the iterative conjugate gradient algorithm is publicly available \cite{kim2018}.  However, in this work, we observe that the basic Landweber iteration algorithm (a form of gradient descent that has guaranteed convergence characteristics for linear least-squares problems \cite{vogel2002}) is easier to adapt to the deep learning formalism used by LORAKI.  Starting from some initial guess $\mathbf{d}_{rec}^{(0)}$, the Landweber iteration procedure corresponding to AC-LORAKS is equivalent to iterating the following equation
\begin{equation}
\mathbf{d}_{rec}^{(i+1)} = \mathcal{U}\left((\mathbf{d}_{rec}^{(i)} - \lambda \mathcal{P}^*\left(\mathcal{P}(\mathbf{d}_{rec}^{(i)})\mathbf{N}\mathbf{N}^H\right)\right) + \mathbf{d}_{zp},\label{eq:update}
\end{equation}
where the linear operator $\mathcal{U}: \mathbb{C}^{N_1 \times N_2 \times L} \rightarrow \mathbb{C}^{N_1 \times N_2 \times L}$ is defined by $\mathcal{U}(\mathbf{x}) \triangleq \mathbf{x} - \mathcal{M}(\mathbf{x})$, the operator $\mathcal{P}^*(\cdot): \mathbb{C}^{P\times Q} \rightarrow \mathbb{C}^{N_1 \times N_2 \times L}$ is the adjoint of $\mathcal{P}(\cdot)$, and $\lambda$ is a step size parameter that needs to be chosen suitably small (as defined by the spectral characteristics of the matrix associated with the linear $\mathcal{P}(\cdot)\mathbf{N}$ operation) to ensure the convergence of the iteration to the optimal least squares solution \cite{vogel2002}.  

Importantly, because the LORAKS matrices constructed by $\mathcal{P}(\cdot)$ are convolution-structured, the Landweber iteration procedure from Eq.~\eqref{eq:update} takes the form of a two-layer convolutional RNN without bias terms or activation functions.  Specifically, each iteration can be written as a convolution layer $g_1(\cdot): \mathbb{C}^{N_1 \times N_2 \times L} \rightarrow \mathbb{C}^{N_1 \times N_2 \times C}$ associated with the linear convolutional operator $\mathcal{P}(\cdot)\mathbf{N}$, a second convolution layer $g_2(\cdot): \mathbb{C}^{N_1 \times N_2 \times C} \rightarrow \mathbb{C}^{N_1 \times N_2 \times L}$ associated with the linear convolutional operator $\mathcal{P}^*(\cdot \mathbf{N}^H)$, and a final projection onto data consistency implemented using the $\mathcal{U}$ operator: 
\begin{equation}
\mathbf{d}_{rec}^{(i+1)} = \mathcal{U}\left(\mathbf{d}_{rec}^{(i)} - \lambda g_2(g_1(\mathbf{d}_{rec}^{(i)})) \right) + \mathbf{d}_{zp}.
\end{equation}
This RNN structure of Landweber-based AC-LORAKS is also shown in Fig.~\ref{fig:network_structure}.   

An important difference between AC-LORAKS and GRAPPA is that AC-LORAKS applies directly to the entire undersampled dataset at once, and does not require separate reconstruction of each channel or enumeration and separate treatment of all of the distinct local sampling configurations.  This can simplify the reconstruction procedure, and for example ensures that AC-LORAKS can be substantially easier to use with non-uniform k-space sampling patterns where the number $J$ of local sampling configurations may be large.  In addition, this can also improve the reconstruction of missing samples whose closest neighboring acquired samples may be far away with respect to the size of the convolution kernels used for reconstruction.

While the convolution kernels used in LORAKS could be rectangular like the kernels employed by most other methods (including RAKI and GRAPPA), LORAKS implementations have classically always relied on ellipsoidal convolution kernels \cite{haldar2014}.  An $R_1 \times R_2$ ellipsoidal convolution kernel can be viewed as a special case of a standard $R_1 \times R_2$ rectangular kernel, where the values in the corners of the rectangle (i.e., the region outside the ellipse inscribed within the rectangle)  are forced to be zero.  Ellipsoidal kernels have several advantages,  including fewer degrees of freedom to achieve the same spatial resolution characteristics as rectangular kernels (e.g., in 2D, an ellipse has $\pi/4 \approx 78.5\%$ of the area of the rectangle that circumscribes it), isotropic resolution characteristics rather than the anisotropic resolution associated with rectangular kernels, and overall better empirical reconstruction performance for a wide range of autoregressive reconstruction methods  \cite{lobos2019}.

\subsection{The Proposed LORAKI Approach}
Inspired by RAKI and AC-LORAKS, LORAKI is implemented by simply including  nonlinear ReLU activation functions within the convolutional RNN architecture of Landweber-based AC-LORAKS.  In particular, starting from an initialization of $\mathbf{d}_{rec}^{(0)} = \mathbf{d}_{zp}$, the LORAKI network iterates the following equation for a total of $K$ iterations:
\begin{equation}
\mathbf{d}_{rec}^{(i+1)} = \mathcal{U}\left(\mathbf{d}_{rec}^{(i)} - \lambda g_2(relu(g_1(\mathbf{d}_{rec}^{(i)}))) \right) + \mathbf{d}_{zp}.
\end{equation}
As before, $g_1(\cdot)$ and $g_2(\cdot)$ are convolution layers without bias terms.  The convolutional RNN structure of LORAKI is also shown in Fig.~\ref{fig:network_structure}. 

Similar to RAKI, the nonlinear structure of LORAKI means that this network cannot be trained with the same relatively simple training procedure used by AC-LORAKS.  Instead, LORAKI can be trained by applying backpropagation and using ACS training data.

As with RAKI, we implement LORAKI using a real-valued deep learning architecture that separates the real and imaginary parts of the data and doubles the effective number of channels.  For ease of implementation, we rely on virtual conjugate coils as described in Refs.~\cite{haldar2017a,kim2018, kim2018c} to capture the LORAKS phase constraints, which doubles again the number of channels. Therefore, we effectively have $4L$ channels, with $\mathbf{d}_{rec}^{(i)} \in \mathbb{R}^{N_1 \times N_2 \times 4L}$,  $g_1(\cdot): \mathbb{R}^{N_1 \times N_2 \times 4L} \rightarrow \mathbb{R}^{N_1 \times N_2 \times C}$, and $g_2(\cdot): \mathbb{C}^{N_1 \times N_2 \times C} \rightarrow \mathbb{C}^{N_1 \times N_2 \times 4L}$, where $C$ is a user-selected number of intermediate channels. To maintain consistency with LORAKS implementations and because it leads to improved empirical performance, we also use ellipsoidal convolution kernels in our implementation of LORAKI.

\subsubsection{Training Considerations and Synthetic ACS data}

For training, LORAKI uses the same ACS data as used by the other three methods described above.  Similar to GRAPPA and RAKI, this ACS data is subsampled to generate paired fully-sampled and undersampled training examples.   However, similar to AC-LORAKS but different from GRAPPA and RAKI, LORAKI is easily compatible with non-uniform sampling patterns like random sampling or partial Fourier acquisition, and there is no need to tailor the reconstruction procedure to the specific local sampling configurations that are present in the acquired data.   This means that when constructing paired fully-sampled and undersampled training examples, the undersampling patterns that are used for training do not need to be a close match to the real undersampling pattern that will be reconstructed.

Similar to RAKI, LORAKI is more complicated and has more parameters than AC-LORAKS or GRAPPA.  As a result, it should generally be expected that LORAKI will require more ACS training data than AC-LORAKS does.  However, since acquiring a substantial amount ACS data may reduce experimental efficiency, it would be preferable if the dependence on acquired ACS data could be reduced.  Recently, we have explored an approach for generating synthetic ACS data that worked fairly well in a different context \cite{kim2018c}.  This approach was based on first performing a fast initial reconstruction of the data, and then using that initial reconstruction result as synthetic ACS data to guide the next stage of LORAKS-based image reconstruction.  In this paper, we observe that if only a small amount of ACS data is acquired, we can potentially use the full k-space data obtained by a fast initial AC-LORAKS reconstruction to provide additional synthetic ACS training data to use with LORAKI.  Since the potential value of this synthetic ACS approach is hard to evaluate theoretically, we will instead evaluate it empirically in the following sections.

\section{Methods}

\subsection{Datasets}
We evaluated LORAKI and compared it against other methods by reconstructing retrospectively-undersampled versions of  fully-sampled in vivo human brain datasets from two different contexts.  In one case, the images were T2-weighted and were acquired with 2D Fourier encoding with a $256\times 187$ acquisition matrix (readout $\times$ phase encoding) on a 3T scanner using a 12-channel head coil.  In the other case, the images were T1-weighted and were acquired with 3D Fourier encoding using an MPRAGE sequence on a 3T scanner with a 32 channel head coil.  In this case, we reconstructed the fully-sampled readout dimension and considered representative 2D slices with acquisition matrix $208 \times 256$ (phase encoding 1 $\times$ phase encoding 2).  Due to the large number of channels, the 32-channel data was coil-compressed to 8 channels prior to reconstruction.  For both datasets, we performed evaluations for 5 representative slices of the same subject.

\subsection{Retrospective Undersampling and Comparisons}

We performed several types of retrospective undersampling experiments.  In one set of experiments, we considered uniform 1D undersampling of the T2-weighted data along the phase encoding dimension, and compared LORAKI (with either real or synthetic ACS data) against GRAPPA, RAKI and AC-LORAKS.  The reconstruction results were evaluated subjectively using visual inspection, as well as quantitatively using standard normalized root-mean-squared error (NRMSE) and structural similarity index (SSIM) error metrics.  For NRMSE, smaller numbers are better with a perfect reconstruction corresponding to an NRMSE value of zero.  For SSIM, larger numbers are better with a perfect reconstruction corresponding to an SSIM value of one.  In one case, we also analyzed the reconstruction error as a function of spatial frequency, using the error spectrum plot (ESP) \cite{kim2018a}.

In another set of experiments, we considered 2D variable-density random undersampling of the T1-weighted data.  Since it is inconvenient to use GRAPPA and RAKI with random sampling, we only compared against AC-LORAKS for this case.  Reconstruction quality was assessed using the same qualitative and quantitative methods used in the previous case.

Further experiments were also performed to evaluate the compatibility of LORAKI with different undersampling patterns, and to assess the impact of different amounts of ACS data and/or synthetic ACS data.

\subsection{Parameter Selection and Optimization}

For concreteness and without loss of generality, we considered an implementation of LORAKI that uses $C=64$ and $K$=5.  We used ellipsoidal convolution kernels with $R_1=R_2=3$ for both convolution layers. Our LORAKI code was implemented in PyTorch.  When generating training data, we used undersampling masks with similar characteristics to the actual data.  I.e., if the data was sampled uniformly, then we also trained the LORAKI network with  uniform sampling examples generated from the ACS data.  All experiments were conducted on Google Colab leveraging an NVidia Tesla K80 GPU.  

The implementation of GRAPPA we compared against used a kernel with $R_1=R_2 = A+1$, where $A$ denotes the acceleration factor of the scan.  This choice implies that, for 1D uniform undersampling patterns, each missing sample is interpolated using the $A+1$ nearest samples from each of the two nearest acquired phase encoding lines.

The implementation of RAKI we compared against used the same choices of network parameters (including kernel sizes, kernel dilation factors, etc.) as described in the original paper \cite{akcakaya2019}.  

The implementation of AC-LORAKS we compared against is publicly available \cite{kim2018}.  We used the ``S''-version of AC-LORAKS (which incorporates support, phase, and parallel imaging constraints), and used ellipsoidal convolution kernels with $R_1=R_2=7$.  The $C$ parameter for AC-LORAKS was optimized on an image-by-image basis, and we report results for the value of $C$ that achieved the smallest NRMSE.  This choice represents a best-case scenario, since the true NRMSE value would not be available for a prospective acquisition.

\section{Results}

\begin{figure}[t]
\centering
\includegraphics{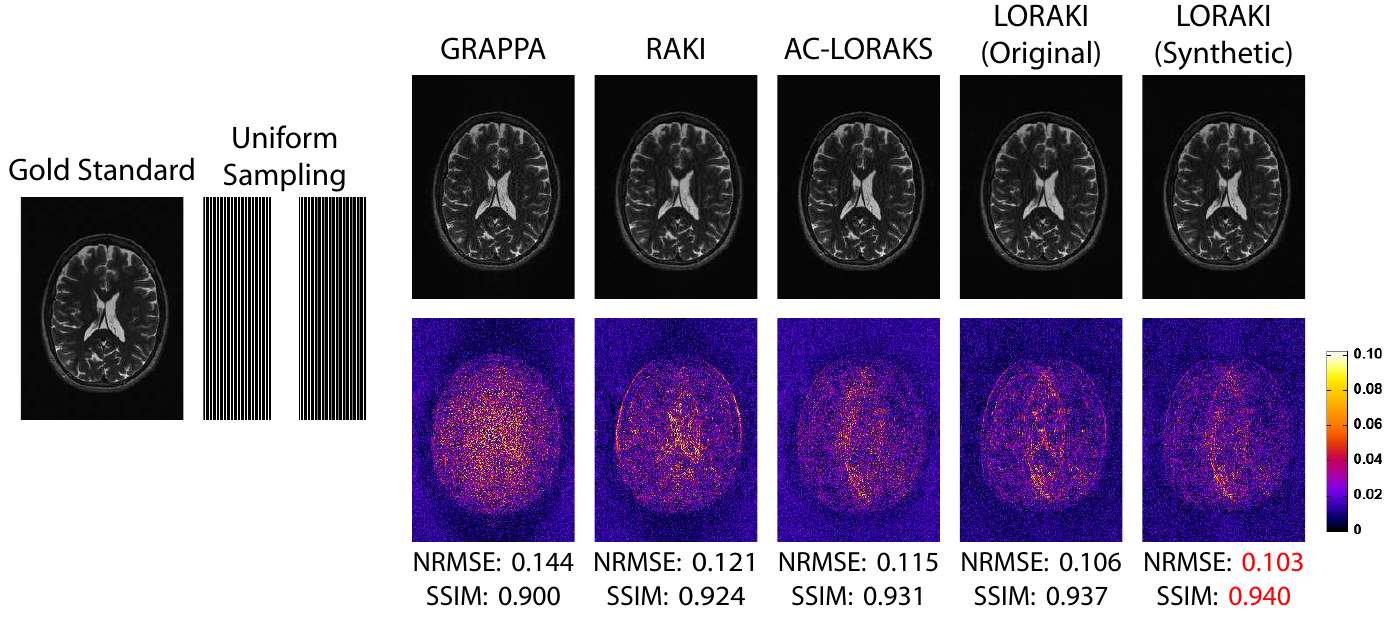} 	
\caption{Representative reconstruction results for uniformly-undersampled T2-weighted data.  The top row shows reconstructed images for one slice in a linear grayscale, where the gold standard image has been normalized to range from 0 (black) to 1 (white). 	The bottom row shows error images with the indicated colorscale.  NRMSE and SSIM values are also shown below each image, with the best values highlighted in red.   }
\label{fig:T2_comparison}
\end{figure}

Figure~\ref{fig:T2_comparison}, Table~\ref{tab:T2_comparison_table}, and supporting Fig.~\ref{fig:supp_esp} show  results from reconstructing uniformly-undersampled T2-weighted data.  In this specific case, we simulated an acquisition that measured every fourth line of k-space, while also fully-acquiring the central 32 phase encoding lines to be used as ACS data.  Taken together, this results in an effective acceleration factor of $2.6\times$.   As can be observed from Fig.~\ref{fig:T2_comparison} and Table~\ref{tab:T2_comparison_table}, the proposed LORAKI approach had the best performance in all cases, with uniformly lower NRMSE values and uniformly higher SSIM values compared to GRAPPA, RAKI, or AC-LORAKS.  We observed similar NRMSE and SSIM values for LORAKI when using the original ACS data or when using synthetic ACS data generated from an initial AC-LORAKS reconstruction of the data.  In this specific case, the amount of acquired ACS data is already relatively high, which may explain the relative lack of impact from using synthetic ACS data.  A corresponding ESP plot shown in supporting Fig.~\ref{fig:supp_esp} shows that LORAKI approaches have consistently similar or better error characteristics than other methods across all spatial frequencies, with the most significant advantage at high-spatial frequencies. 

\begin{figure}[t]
\centering
\includegraphics{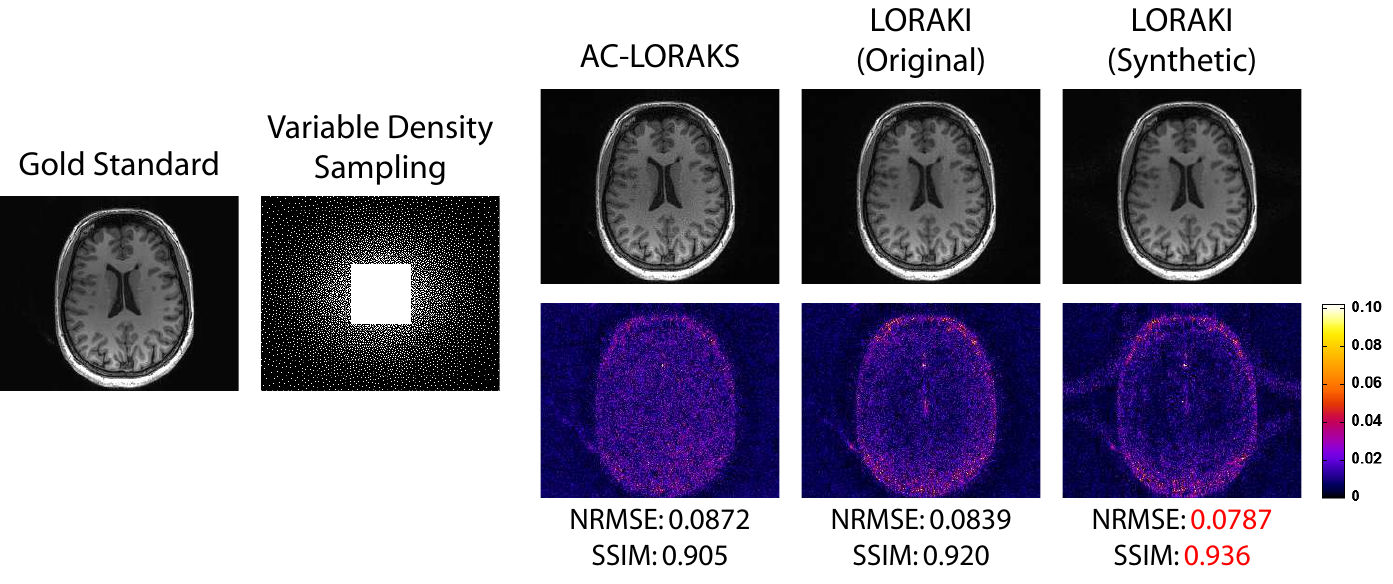}
\caption{Representative reconstruction results for randomly-undersampled T1-weighted data.  The top row shows reconstructed images for one slice in a linear grayscale, where the gold standard image has been normalized to range from 0 (black) to 1 (white). 	The bottom row shows error images with the indicated colorscale.  NRMSE and SSIM values are also shown below each image, with the best values highlighted in red.}
\label{fig:MPRAGE_comparison}
\end{figure}

Figure~\ref{fig:MPRAGE_comparison}, Table~\ref{tab:MPRAGE_comparison_table}, and supporting Fig.~\ref{fig:supp_esp} show results from reconstructing randomly-undersampled T1-weighted data.  In this specific case, we simulated an acquisition with an effective acceleration factor of $5.2\times$ (including samples from a fully-sampled $64\times 64$ ACS region at the center of k-space). Similar to the previous case, LORAKI  had uniformly smaller NRMSE and larger SSIM values compared to AC-LORAKS, with the most significant error improvements at high spatial frequencies.  GRAPPA and RAKI reconstruction were not performed in this case, due to the large number of local sampling configurations resulting from random sampling.     As before, there was not a big difference between using the original ACS data versus using synthetic ACS data, which might be explained by the relatively large size of the acquired ACS data.

\begin{figure}[t]
\centering
\includegraphics{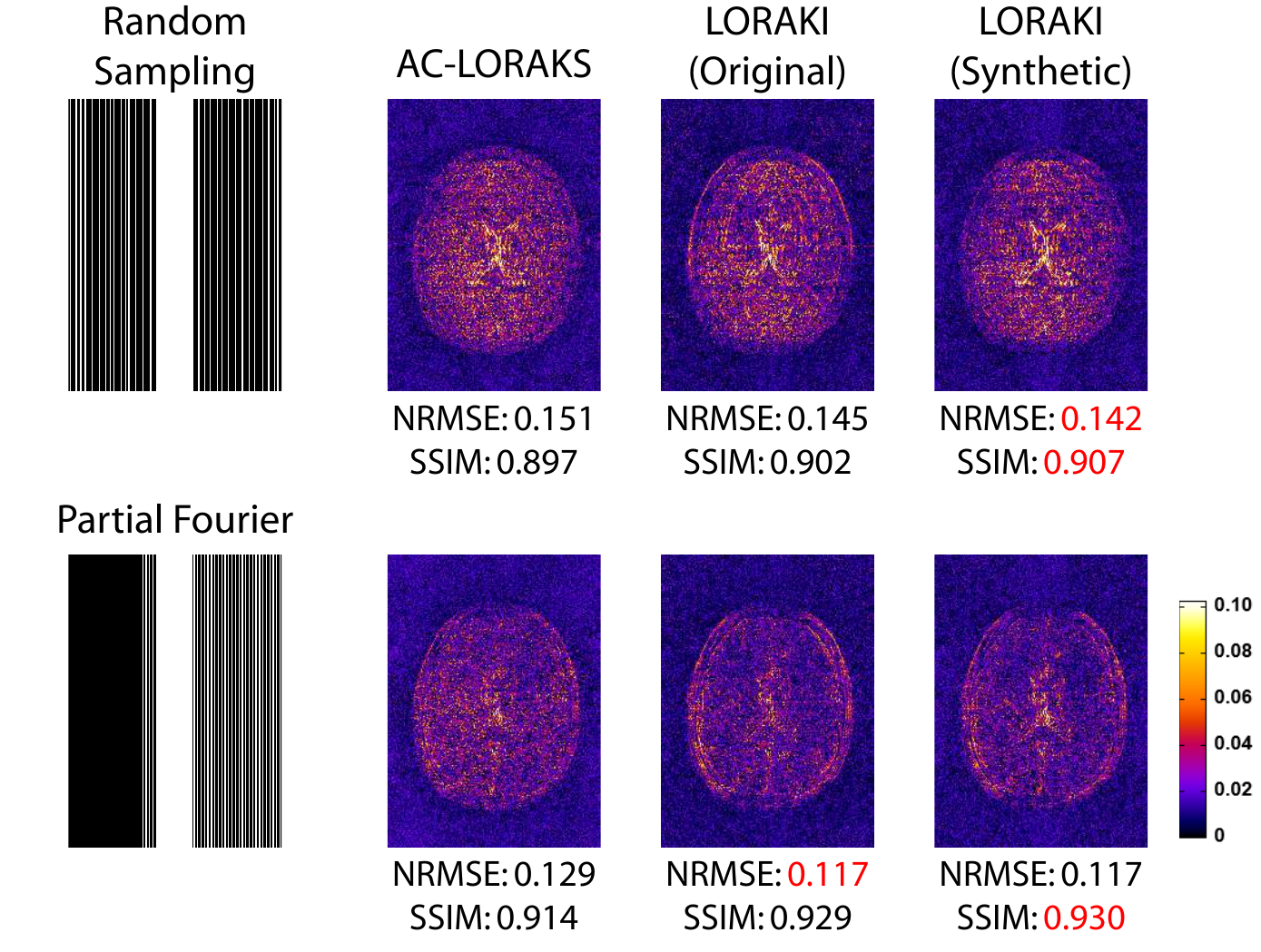} 	
\caption{AC-LORAKS and LORAKI reconstruction results for T2-weighted data with different non-uniform sampling patterns. (top) Random sampling.  (bottom) Partial Fourier sampling.  Error images are shown using the same colorscale from Fig.~2. } 
\label{fig:various_sampling}
\end{figure}

To evaluate the hypothesis that LORAKI would be compatible with a range of different sampling patterns (a characteristic that it should inherit from AC-LORAKS), we performed reconstruction of the T2-weighted data from random undersampling (an effective acceleration factor of $3\times$, including  32 fully-sampled lines  of central k-space to be used as ACS data) and partial Fourier undersampling (5/8ths partial Fourier sampling  including  32 fully-sampled lines  of central k-space to be used as ACS data,  with uniform sampling of the remaining k-space resulting in an effective acceleration factor of $3\times$).  LORAKI reconstruction results for one slice are shown in Fig.~\ref{fig:various_sampling}, and AC-LORAKS results are also included for reference.  As can be seen, the advantage of LORAKI over AC-LORAKS is still observed for these sampling patterns, and there is still not a major difference between using the original ACS data and synthetic ACS data.  

\begin{figure}[t]
\centering
\includegraphics{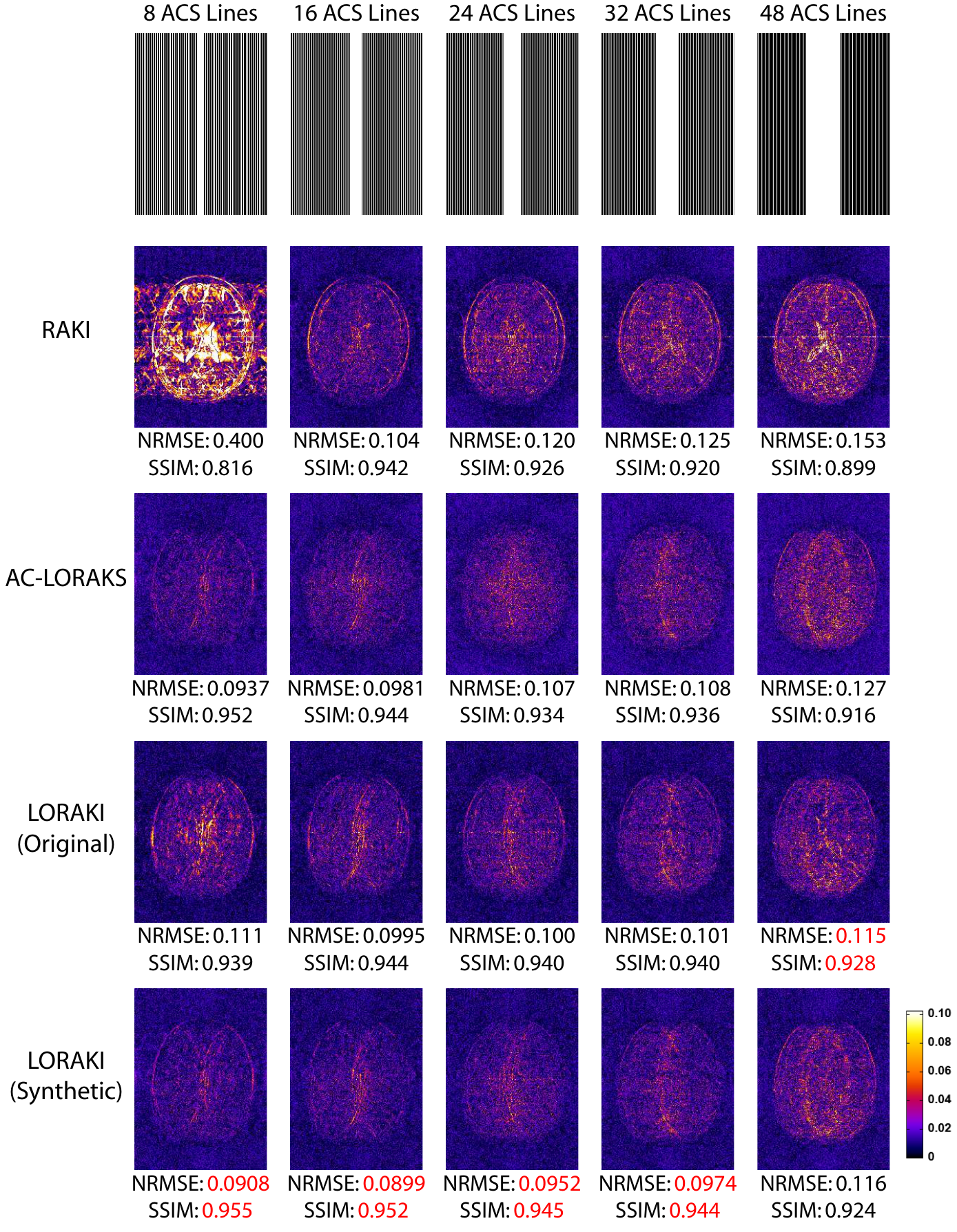} 	
\caption{Reconstruction results for T2-weighted data with varying amounts of ACS data. Error images are shown using the same colorscale from Fig.~2.}
\label{fig:ACS_size_T2} 
\end{figure}

\begin{figure}[t]
\centering
\includegraphics{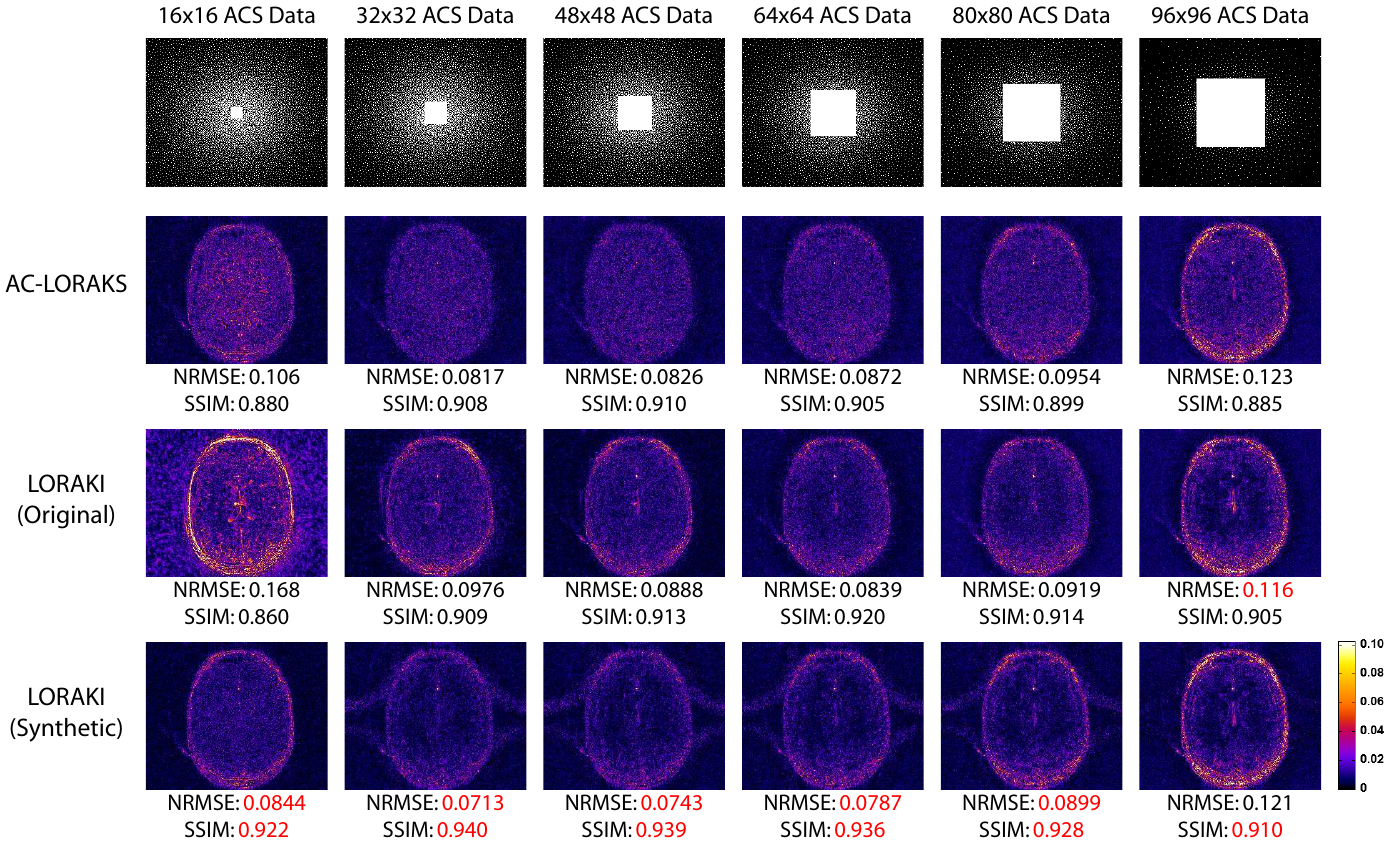} 	
\caption{Reconstruction results for T1-weighted data with varying amounts of ACS data. Error images are shown using the same colorscale from Fig.~3. }
\label{fig:ACS_size_MPRAGE}
\end{figure}

All of the previous examples used a relatively large amount of acquired ACS data, which is likely beneficial for methods like RAKI and LORAKI, but which might also reduce experimental efficiency.  In the next set of experiments, we performed reconstructions with different amounts of ACS data, while holding the effective acceleration factor fixed.  For the T2-weighted data, we varied the number of fully-sampled lines at the center of k-space, and performed uniform undersampling of the remainder of k-space.  The sample spacing was adjusted so that the total number of lines was equal for each case, with an effective acceleration factor of 2.5$\times$.    For the T1-weighted data, we varied the size of the fully-sampled region at the center of k-space, and used variable density random sampling for the remainder of k-space.  The total number of samples was held fixed in each case, with an effective acceleration factor of 5.2$\times$.  Results for T2-weighted and T1-weighted datasets are shown in Figs.~\ref{fig:ACS_size_T2} and \ref{fig:ACS_size_MPRAGE}, respectively.  As expected, LORAKI using the original ACS data works better than AC-LORAKS when the amount of ACS data is large, but AC-LORAKS yields better results when the amount of ACS data is small.  However, these results also demonstrate that synthetic ACS data is potentially quite valuable when the amount of actual ACS data is relatively small.  In particular, LORAKI with synthetic ACS data appears to consistently outperform AC-LORAKS across all cases.  Notably, we also observe that both LORAKI and AC-LORAKS appear to consistently outperform RAKI in Fig.~\ref{fig:ACS_size_T2}.

\section{Discussion}

\begin{figure}[t]
\centering
\includegraphics{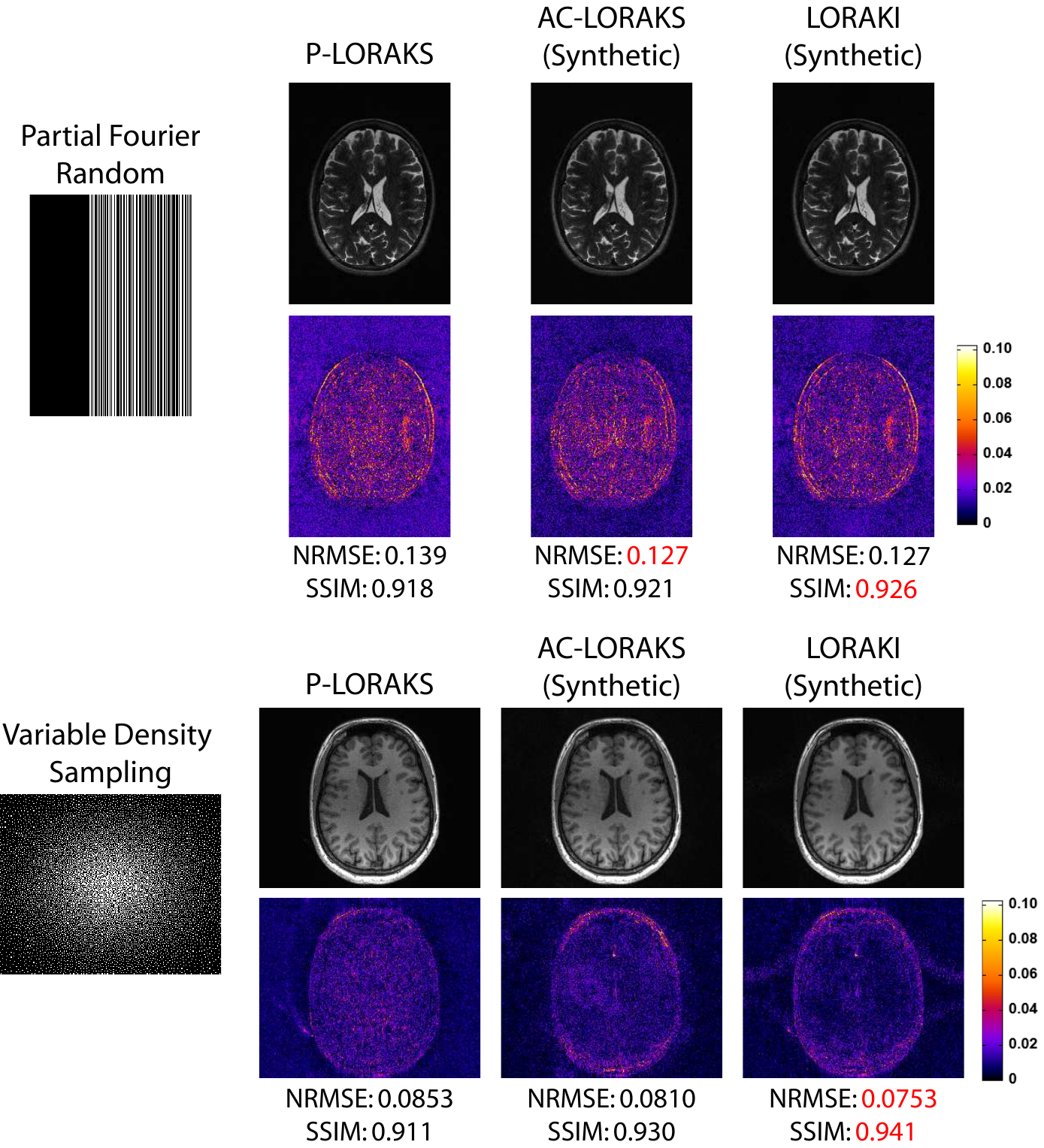}
\caption{Evaluation of calibrationless reconstruction using synthetic ACS data.  Error images are shown using the same colorscales from Figs. 2 and 3. }
\label{fig:calibrationless}
\end{figure}

The results shown in the previous section demonstrated that LORAKI has potential advantages compared to existing AA methods when sufficient ACS data is available, and also that synthetic ACS training data is potentially useful for scenarios where it may be impractical to acquire a large amount of actual ACS data.  In practice, there can also be certain scenarios where no ACS training data is available, where existing calibrationless reconstruction methods  like SAKE and LORAKS (which are also based on linear autoregressive modeling principles) have previously demonstrated value  \cite{shin2014, haldar2014,haldar2016}.  While the LORAKI formulation does not directly address calibrationless scenarios, it is worth noting that LORAKI could also potentially be applied to such scenarios if synthetic ACS data can be generated (e.g., by applying a calibrationless reconstruction method as an initial step).  As an initial proof-of-principle for this idea, we performed two different calibrationless simulations, as shown in Fig.~\ref{fig:calibrationless}.  With the T2-weighted data, we simulated calibrationless random partial Fourier undersampling with an effective acceleration factor of $3.5\times$.  With the T1-weighted data, we simulated calibrationless variable density random sampling with an effective acceleration factor of $5\times$.  In both cases, we used the ``S''-version of the nonconvex P-LORAKS method \cite{haldar2016} (using publicly available software \cite{kim2018a}) to generate an initial reconstruction.  This initial reconstruction was then used as synthetic ACS training data to train LORAKI, and LORAKI reconstruction was then performed.  Reconstruction results are shown in Fig.~\ref{fig:calibrationless}, and we also show the P-LORAKS reconstructions and AC-LORAKS reconstructions (trained using the P-LORAKS reconstruction as ACS data) for reference.  As can be seen, the LORAKI reconstruction frequently has the best performance metrics compared to P-LORAKS and AC-LORAKS. The one exception is that AC-LORAKS has a slightly better NRMSE value for the T2-weighted data, although the difference in NRMSE between LORAKI and AC-LORAKS is nearly negligible in this case (i.e., an NRMSE of 0.1274 for LORAKI versus 0.1271 for AC-LORAKS).  These results confirm that LORAKI-type approaches can still have relevance to calibrationless scenarios.

\begin{figure}[t]
\centering
\includegraphics{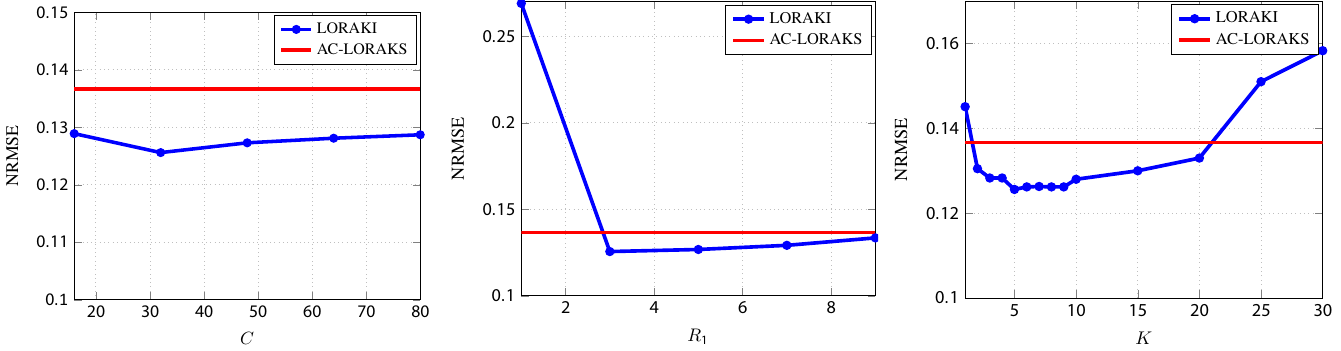}
\caption{Evaluating the effects of different LORAKI network parameters  on reconstruction performance.  The first plot shows the effects of varying the number of hidden-channel layers $C$ while holding the kernel size fixed at $R_1=R_2=3$ and the number of iterations fixed at $K=5$.  The second plots shows the effects of varying $R_1$, while setting $R_2=R_1$ and holding the other parameters fixed at $C=64$ and $K=5$.  The final plot shows the effects of varying $K$, while holding the other parameters fixed at $C=64$ and $R_1=R_2=3$. For reference, the NRMSE value for AC-LORAKS reconstruction with optimized parameters is also shown (the AC-LORAKS parameters are not varied in this plot). }
\label{fig:param_tuning}
\end{figure}

The results shown in this paper were all generated using the same set of LORAKI network parameters.  However, different choices of these parameters are expected to have an impact on reconstruction performance.  Figure~\ref{fig:param_tuning} illustrates the impact of the parameters $C$ (the number of channels in the hidden layer), $K$ (the number of RNN iterations), and $R_1 \times R_2$ (the size of the ellipsoidal convolution kernels) on reconstruction performance.  The results in this figure were generated based on reconstructing the uniformly undersampled T2-weighted data, with 64 ACS lines and an effective acceleration factor of $3\times$.  For reference, the NRMSE for AC-LORAKS (with optimized AC-LORAKS parameters) is also shown.  As can be seen, the performance of LORAKI appears to be relatively robust with respect to variations in these parameters, and maintains an advantage over AC-LORAKS across a wide range of parameter settings.

Unlike most other deep learning methods, the training procedure for RAKI and LORAKI is scan-specific and therefore must be performed online.  As a result, training time becomes an important consideration.  For the results we've shown, network training required approximately 1 hour on Google Colab for LORAKI.    While this training time is relatively long, we should note that this implementation was designed for simple proof-of-principle evaluation, and we did not spend much effort in optimizing training speed. Substantial speedups may be possible from using optimized hardware and more efficient training algorithms.

In addition, while we've demonstrated that LORAKI improves reconstruction performance over similar AA methods, we should also mention that the LORAKI network is still relatively simple at this stage, and we have not yet considered the use of more advanced deep learning strategies such as dropout \cite{srivastava2014} or batch normalization \cite{loffe2015}.  Combination of LORAKI-type ideas with other forms of constrained reconstruction may also prove to be a fruitful direction for future work.

\section{Conclusion}
This work introduced LORAKI, a novel scan-specific autocalibrated RNN approach for nonlinear autoregressive MRI reconstruction in k-space that was motivated by ideas from previous RAKI and AC-LORAKS methods.  LORAKI is designed to automatically capture the same support, phase, and parallel imaging constraints as AC-LORAKS while also maintaining compatibility with a wide range of k-space sampling patterns.  However, different from AC-LORAKS but similar to RAKI, the reconstruction procedure in LORAKI is nonlinear, and capable of capturing more complicated autoregressive relationships.   Our evaluations with retrospectively undersampled MRI data suggest that LORAKI can outperform similar existing reconstruction methods in many situations, and we envision that the further development of this kind of approach may enable even bigger gains in the future.

\bibliography{reference}

\newpage 

\clearpage 

\clearpage 

\begin{table}[t]
\centering
\includegraphics{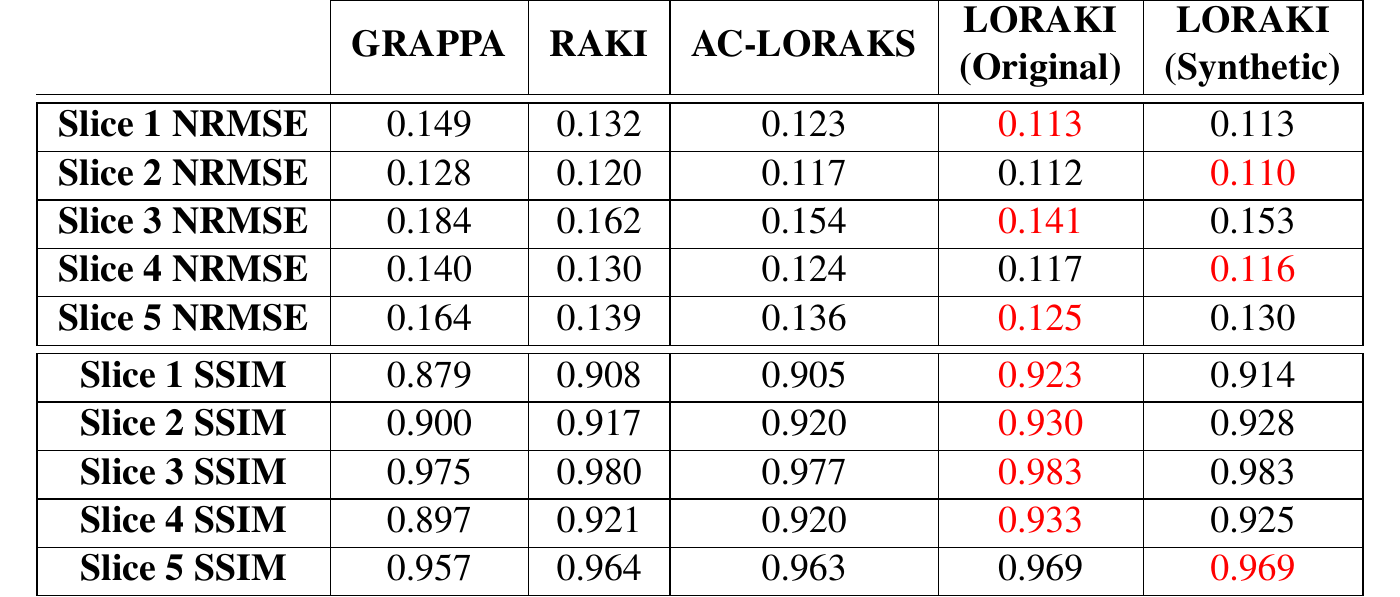} 	
\caption{Quantitative reconstruction performance metrics for 5 different slices of the T2-weighted dataset, using the same sampling pattern shown in Fig.~2.  The best performance metrics are highlighted in red.}
\label{tab:T2_comparison_table}
\end{table}

\clearpage 

\clearpage 

\begin{table}[t]
\centering
\includegraphics{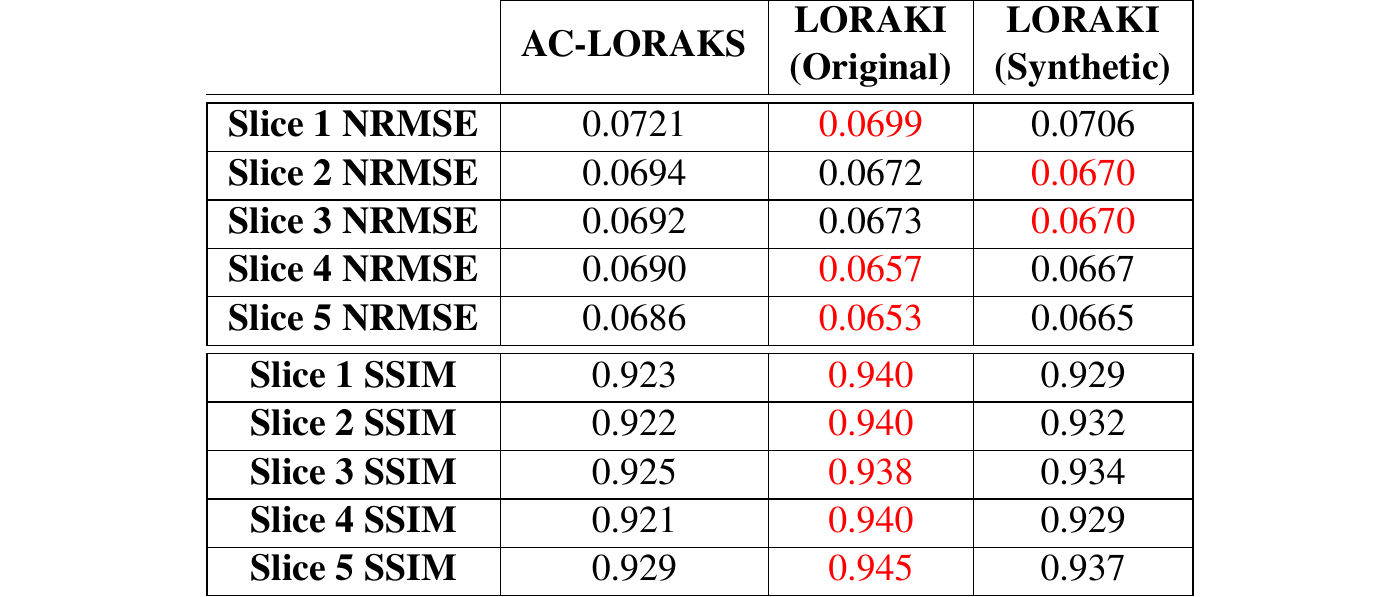} 	
\caption{Quantitative reconstruction performance metrics for 5 different slices of the T1-weighted dataset, using the same sampling pattern shown in Fig.~3.    The best performance metrics are highlighted in red. }
\label{tab:MPRAGE_comparison_table}
\end{table}

\clearpage 

\clearpage

\clearpage 

\clearpage 

\clearpage 

\clearpage
\setcounter{figure}{0} 
\renewcommand{\thefigure}{S\arabic{figure}}

\begin{figure}
	\centering
	\includegraphics{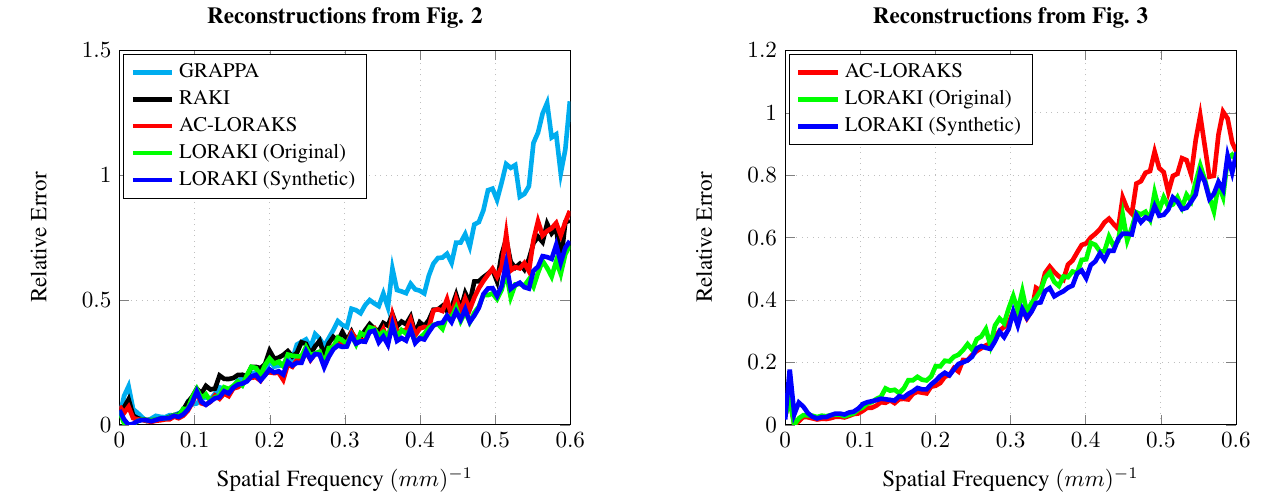}
	\caption{Error spectrum plots \cite{kim2018a} corresponding to (left) the T2-weighted reconstruction results shown in Fig.~2 and (right) the T1-weighted reconstruction results shown in Fig.~3. }
	\label{fig:supp_esp}
\end{figure}

\end{document}